\documentstyle[prl,aps,psfig,twocolumn]{revtex}

\begin{document}

\title{Collective excitations of Bose-Einstein condensed
gases at finite temperatures}

\author{R.\ J.\ Dodd~\cite{NIST}}
\address{Institute for Physical Science and Technology,
University of Maryland at College Park,
College Park, MD 20742.}

\author{Mark Edwards\cite{NIST}}
\address{Department of Physics,
Georgia Southern University,
Statesboro, GA 30460-8031.}

\author{Charles W.\ Clark}
\address{Electron and Optical Physics Division,
Physics Laboratory,
National Institute of Standards and Technology,\\
Technology Administration,
U.\ S.\ Department of Commerce,
Gaithersburg, MD 20899-0001.}

\author{K.\ Burnett~\cite{NIST}}
\address{Clarendon Laboratory, Department of Physics,
University of Oxford,
Parks Road, Oxford OX1 3PU, United Kingdom.}

\date{\today}

\maketitle

\begin{abstract}
We have applied the Popov version of the Hartree-Fock-Bogoliubov
(HFB) approximation to calculate the finite-temperature excitation
spectrum of a Bose-Einstein condensate (BEC) of $^{87}$Rb atoms.
For lower values of the temperature, we find excellent agreement
with recently-published experimental data for the JILA TOP trap.
In contrast to recent comparison of the results of HFB--Popov theory
with experimental condensate fractions and specific heats, there is
disagreement of the theoretical and recent experimental results
near the BEC phase transition temperature.
\end{abstract}

\draft

Laboratory realizations of gaseous Bose-Einstein condensates
(BECs)~\cite{BEC!,HULET,KETTERLE} have prompted vigorous
experimental~\cite{EXP_BEC_PROPERTIES} investigations of the
temperature-dependent properties of these mesoscopic quantum
systems.  The current theoretical interest in such condensates
derives, in part, from the fact that experimental tests of
many--body theories that are thought to apply to BEC --- a
phenomenon occuring in many areas of physics --- can be performed
for the first time.  An accurate theory of such systems is therefore
of fundamental interest, and will also have practical applications.
In this Letter, we explore the limits of validity of the simplest
temperature--dependent mean-field theory --- a simplified version
of the Hartree--Fock--Bogoliubov (HFB) approximation originally
introduced by Popov~\CITE{POPOV} --- by presenting the first
comparison with experiment of this theory's predictions for
condensate excitation spectra at $T > 0$.

Condensate properties predicted by such theories include condensate
and thermal--atom spatial density profiles, condensate fractions,
specific heats, and excitation frequencies.  Of these properties,
excitation spectra provide the most sensitive test of the applicability
of competing theories since the other quantities listed
depend on sums over states and are thus insensitive to small errors
in the excitation spectrum.  One example of this can be found in the
approach of zero--temperature excitation frequencies to the
Thomas--Fermi limit (i.e., the limit $N_0\rightarrow\infty$, where
$N_0$ is the number of condensate atoms) as $N_0$ increases.  For
$^{87}$Rb condensates confined in the JILA TOP (time--averaged
orbiting potential) trap where $N_0 > 4000$ atoms, the Thomas--Fermi
predictions for spatial density profiles are extremely accurate while
low--order excitation frequencies can differ by as much as 10\%.

There is good {\em a priori} reason to expect  that the HFB--Popov
theory should provide good predictions of experimental $T$--dependent
collective excitation frequencies.  Measurements of zero--temperature
frequencies~\cite{T0_EXC_EXP} exhibited excellent agreement with the
predictions of zero-temperature, mean field theory~\cite{T0_EXC_THY}.
Furthermore, semiclassical variants of the HFB--Popov theory have exhibited
excellent agreement with experiment for $T$--dependent condensate
fractions and specific heats for temperatures up to near $T_c$~\cite
{Popov_compare}.  The HFB--Popov theory is a finite-temperature
extension of mean-field theory which provides self-consistent treatment
of the condensed and thermal components of the gas and which should
describe the linear response of the condensate to small-amplitude
mechanical disturbances~\cite{GRIFFIN}.

Although the HFB-Popov equations have been derived elsewhere~\cite
{GRIFFIN,FETTER}, we shall briefly state the physics behind the basic
equations here.  The confined Bose gas is portrayed as a thermodynamic
equilibrium system under the grand canonical ensemble whose
thermodynamic variables are $N$, the total number of trapped atoms, $T$,
the absolute temperature, and either $N_{0}$ or $\mu$, the chemical
potential.  The system hamiltonian has the form
\begin{eqnarray}
K  \equiv H-\mu N &=& \int\;d{\bf r} \hat{\psi}^{\dagger}({\bf r})
(H_{0}-\mu) \hat{\psi}({\bf r}) \nonumber \\
&+& \frac{U_{0}}{2} \int\;d{\bf r} \hat{\psi}^{\dagger}({\bf r})
\hat{\psi}^{\dagger}({\bf r})\hat{\psi}({\bf r})\hat{\psi}({\bf r}) \;\;.
\label{gch}
\end{eqnarray}
where $\hat{\psi}({\bf r})$ is the Bose field operator which annhililates
an atom at position ${\bf r}$ and $H_{0} = \frac{-\hbar^{2}}{2 M} \nabla^{2}
+ V_{\rm trap}({\bf r})$ is the bare trap Hamiltonian.  For the system
treated here, the trap potential $V_{\rm trap}({\bf r}) = M \left(
\omega_{\rho}^{2}\rho^{2} + \omega_{z}^{2}z^{2}\right)/2$, where $M$ is
the atomic mass, and $\omega_{\rho}$ and $\omega_{z}$ are the radial and
axial trap frequencies, respectively.  The quantity $U_{0} = 4\pi\hbar^{2}
a/M$ is a measure of the interaction strength between atoms, with $a$ being
the scattering length for zero--energy binary atomic collisions, taken to
be $109a_0$ for $^{87}$Rb~\cite{Heinzen}, where $a_0$ is the Bohr radius.

The Bose field operator is decomposed into a $c$--number condensate
wave function plus an operator describing the non--condensate part:
$\hat{\psi}({\bf r}) = N_{0}^{1/2}\phi({\bf r}) + \tilde{\psi}({\bf r})$
and inserted into Eq.\ (\ref{gch}).  When terms cubic and quartic
in $\tilde{\psi}({\bf r})$ are treated within the mean--field
approximation the grand--canonical hamiltonian reduces to a sum of
four terms: $K = K_0 + K_1 + K_1^{\dag} + K_2$.  The first term,
$K_0$ is a $c$--number, the second and third terms are linear
in $\tilde{\psi}({\bf r})$ and $\tilde{\psi}^{\dag}({\bf r})$ and
the last term is quadratic in these quantities.  It is easy to show
that the linear terms vanishes identically if $\phi({\bf r})$ satisfies
the generalized Gross-Pitaevskii (GP) equation
\begin{equation}
\left\{H_{0} + U_{0} \left[ N_{0}\left|\phi({\bf r})\right|^{2} +
2\tilde{n}({\bf r})\right] \right\} \phi({\bf r}) = \mu \phi({\bf r}) ,
\label{gen_gp_eq}
\end{equation}
where $\tilde{n}({\bf r})$ is the density of non--condensate atoms.
Note that the condensate wavefunction is normalized to unity.

The term $K_2$ has the form
\begin{eqnarray}
K_2 &=& \int\;d{\bf r} \tilde{\psi}^{\dagger}({\bf r}) {\cal L}
\tilde{\psi}({\bf r})  +
\frac{N_{0}U_{0}}{2}\int\;d{\bf r} \left(\phi\right)^2({\bf r})
\tilde{\psi}^{\dagger}({\bf r}) \tilde{\psi}^{\dagger}({\bf r})
\nonumber \\
&+& \frac{N_{0}U_{0}}{2}\int\;d{\bf r} \left(\phi^{\ast}\right)^{2}({\bf r})
\tilde{\psi}({\bf r})\tilde{\psi}({\bf r}) \;\;.
\label{K2}
\end{eqnarray}
where ${\cal L} \equiv H_{0} + 2U_{0}n({\bf r}) - \mu$ and $n({\bf r})
= N_{0}\left|\phi({\bf r})\right|^{2} + \tilde{n}({\bf r})$ is the
total trapped-atom density.  The term $K_2$ can be diagonalized by the
following Bogoliubov transformation
\begin{eqnarray}
\tilde{\psi}({\bf r}) &=& \sum_j ( u_j({\bf r}) \alpha_j + v_j^{\ast}({\bf r})
\alpha_j^{\dagger}) \;\;, \nonumber \\
\tilde{\psi}^{\dagger}({\bf r}) &=& \sum_j (u_j^{\ast}({\bf r})
\alpha_j^{\dagger} + v_j({\bf r})\alpha_j) \;\;.
\label{trans}
\end{eqnarray}
if the quasi-particle amplitudes $u_{j}({\bf r})$ and $v_{j}({\bf r})$,
satisfy the coupled HFB-Popov equations:
\begin{eqnarray}
{\cal L} u_j({\bf r}) +
N_{0}U_{0}\left|\phi({\bf r})\right|^{2}v_j({\bf r}) &=& E_j
u_j({\bf r})
\nonumber \\
{\cal L} v_j({\bf r}) +
N_{0}U_{0}\left|\phi({\bf r})\right|^{2}u_j({\bf r}) &=& -E_j
v_j({\bf r})\,,
\label{HFB}
\end{eqnarray}
The $\alpha_{j}$ and $\alpha_{j}^{\dag}$ are quasi--particle
annihilation and creation operators that satisfy the usual Bose
commutation relations.

The density of the thermal component of the gas $\tilde{n}({\bf r})
\equiv \langle\tilde{\psi}^{\dag}\tilde{\psi}\rangle$ and thus can be
written in terms of the quasi--particle amplitudes as
\begin{equation}
\tilde{n}({\bf r}) = \sum_{j}\left\{\left[\left|u_{j}({\bf r})\right|^{2}
+ \left|v_{j}({\bf r})\right|^{2}\right]N_{j} +
\left|v_{j}({\bf r})\right|^{2}\right\},
\label{nc_density}
\end{equation}
where $N_{j} = \left(e^{\beta E_{j}} - 1\right)^{-1}$ is the
Bose-Einstein factor, and $\beta = \left(k_{\rm B}T\right)^{-1}$
with $k_{\rm B}$ the Boltzmann constant.  The total number of trapped
atoms, $N$, is given by
\begin{equation}
N = \int d{\bf r}~ n({\bf r}) = N_{0} +
\int d{\bf r}~\tilde{n}({\bf r}).
\label{total_n}
\end{equation}
Our version of Eq.\ (\ref{HFB}) differs from that of Ref.\ \cite{HUTCHINSON}
via a sign change in the definition of $v_j({\bf r})$.

Equations (\ref{gen_gp_eq}), (\ref{HFB}), (\ref{nc_density}), and
(\ref{total_n}) form a closed system of equations that we have referred
to as the ``HFB--Popov'' equations.  We have numerically solved these
equations under conditions appropriate to $^{87}$Rb atoms confined in
the JILA TOP trap~\cite{T_DEPENDENT_EXC}.  We choose our state variables
to be $\{T,\mu,N\}$, fix $T$ and $\mu$, and then determine $N$ by
solving the HFB--Popov equations.  This is equivalent to the
alternative triple of state variables $\{T,N_0,N\}$, since there is a
one-to-one relationship between $N_{0}$ and $\mu$.

We have solved the HFB--Popov equations by an iterative procedure,
each cycle of the iteration consisting of two steps.  In the first step
of each cycle, we solve Eq.\ (\ref{gen_gp_eq}) for new values of
$\phi({\bf r})$ and $N_0$ with a basis--set approach as described
previously~\cite{EDCRB96} using $\tilde{n}({\bf r})$ obtained in the
previous cycle.  In the second step, we solve Eqs.\ (\ref{HFB})
using $\tilde{n}({\bf r})$ from the previous cycle, and the newly generated
values of $\phi({\bf r})$ and $N_0$.  With the quasi-particle amplitudes
expanded in the trap basis, Eqs.\ (\ref{HFB}) yield a generalized
matrix eigenvalue problem for the basis--set coefficients.  We recast
the generalized matrix eigenvalue problem by using a decoupling
transformation consisting of taking the sum and difference of Eqs.\ %
(\ref{HFB}).  This transformation is equivalent to that of Hutchinson
{\em et al.}~\cite{HUTCHINSON}, except that it is expressed in terms
of basis set expansion coefficients.  Completion of this step
yields the $\left\{u_{j}({\bf r}),v_{j}({\bf r})\right\}$, and $E_{j}$
which are used in Eq.\ (\ref{nc_density}) to update $\tilde{n}({\bf r})$.
Equation (\ref{total_n}) then updates the total number of trapped
atoms, $N$.  Convergence is reached when the change in $N$ from one
cycle to the next is smaller than a specified tolerance.  To obtain
converged results at high-temperatures, we add a correction to the total
number of atoms, $N$, at each iteration cycle.  High-energy
quasi--particle eigenfunctions have negligible overlap with the
condensate wave function, so their presence in the thermal sum
of Eq.(\ref{nc_density}) does not significantly modify the low-lying
excitation frequencies, but does contribute to the value of $N$.

We have checked the accuracy of our numerical work by writing two
independent codes, which produce identical answers.  The ideal gas
result is recovered when we set $a = 0$, and we have reproduced the
results reported in Ref.~\cite{HUTCHINSON}.  We now discuss the
comparison of this approach with experiment.

\begin{figure}[!h]
\begin{center}\
\psfig{file=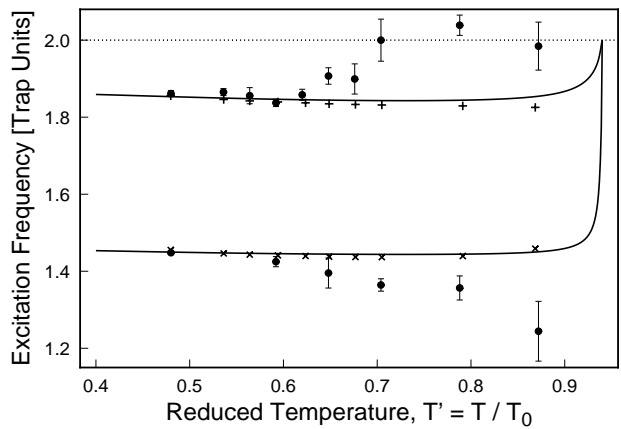,width=3.4in}
\end{center}
\caption{The experimental, temperature dependent excitation
spectrum in the JILA TOP trap (filled
circles) versus the HFB-Popov predictions for the $m=0$ mode (top,
labeled by ``$+$'') and the $m=2$ mode (bottom, labeled by
``$\times$'').  The solid curves are excitation frequencies for a {\em
zero-temperature} condensate having the same number of condensate
atoms as the experimental condensate in the finite--$T$ cloud.}
\label{fig1}
\end{figure}

\begin{figure}[!h]
\begin{center}\
\psfig{file=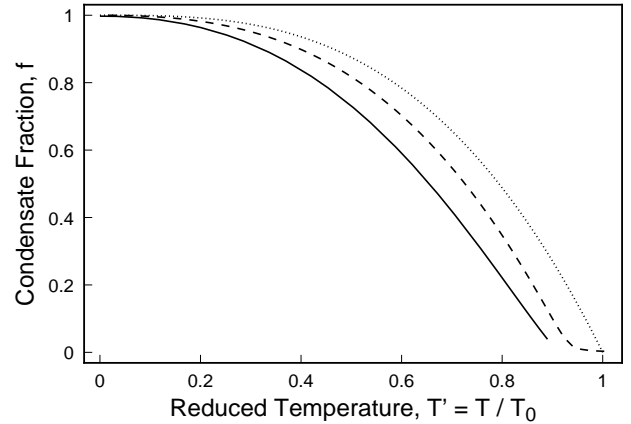,width=3.4in}
\end{center}
\caption{A plot of the condensate fraction as a function of
$T^{\prime}$  (solid curve) for the JILA TOP trap in which $N$ is
fixed at $2000$ atoms.  The same quantity is shown for the ideal gas in
the thermodynamic limit (dotted curve) and for $2000$ atoms
(dashed curve) for comparison.}
\label{fig2}
\end{figure}

Figure \ref{fig1} compares the experimental \cite{T_DEPENDENT_EXC}
excitation spectrum of $^{87}$Rb in the JILA TOP trap, {\it vs.} our
HFB-Popov results for the $m=0$ and $m = 2$ modes.  The abscissa is
the scaled temperature $T^{\prime} = T/T_{0}(N,\bar{\omega})$ where
$T_{0}(N,\omega)\equiv \hbar\omega/k_{B}\left[N/\zeta(3)\right]^{1/3}$
is the theoretical transition temperature for an ideal, trapped Bose
gas and $\bar{\omega} = \left(\omega_{\rho}^{2}\omega_{z} \right)^{1/3}$.
The ordinate is the excitation frequency expressed in units of
$\omega_{\rho}$.  Our results were obtained using the experimental value
of $T$, and a value of $\mu$ that yielded the experimentally determined
value of $N$.  Thus, as for our previous treatment of zero-temperature
excitation spectra \cite{T0_EXC_THY}, {\em this calculation contains no
adjustable parameters}.  The agreement between theory and experiment is
very good (on the order of $5$\%) for low and intermediate temperatures
($T^{\prime} \le 0.65$).  It should be noted that, as depicted in
Fig. \ref{fig2}, the high end of this temperature range corresponds to a 
non-condensate fraction of about 50\%.  However, as the temperature increases,
the HFB-Popov excitation frequencies diverge from the experimental data.
This feature of the comparison holds true for both $m=0$ and $m=2$ modes.

\begin{figure}[!h]
\begin{center}\
\psfig{file=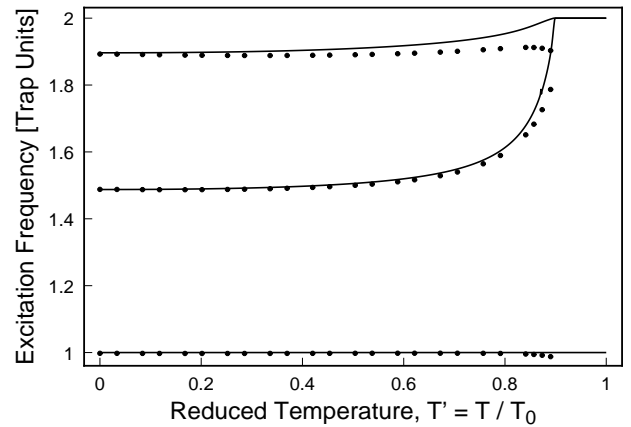,width=3.4in}
\end{center}
\caption{HFB-Popov excitation frequencies (filled circles)
for the $m=0$ (top), $m=2$ (middle), and the $m=1$ modes
(bottom) for a cold-atom cloud having $N = 2000$ atoms.
Overlaid (solid lines) are the frequencies
for a zero-temperature system with the same number
$N_{0}$ of condensate atoms as in the finite-temperature
system.}
\label{fig3}
\end{figure}

The behavior of the calculated excitation frequencies can be
understood in a simple way.  The HFB--Popov equations determine
the equation of state for the state variables $\{N,N_{0},T\}$,
so that, given the values of $N$ and $T$, a unique $N_{0}$
is determined.  For fixed $N$, this
relationship generates the
condensate fraction $N_{0}/N$ as a function of $T$,
which is shown in Fig. \ref{fig2} for the JILA TOP trap
with $N = 2000$.  One can easily predict the temperature-dependent
mode frequencies for the $N = 2000$ system by finding the
number of condensate atoms, $N_{0}$, from Fig.~\ref{fig2},
and then determining the {\em zero-temperature}
excitation
frequency of a condensate with $N_{0}$ atoms, which is
a much simpler calculation.
Figure \ref{fig3} shows a comparison of the three lowest
frequencies numerically determined from the HFB-Popov equations,
and by the equivalent $T=0$ method just described.
The agreement of these two approaches is very
good over nearly the entire temperature range.  The two solid curves in
Fig.~\ref{fig1} are the frequencies determined by the same procedure
except that the number of condensate atoms was taken from experiment.
In short, the principal effect of finite temperature on the
HFB-Popov excitation spectra is largely an effect of
{\it condensate depletion}: the dynamics of the finite-temperature
condensate are essentially the same as those of a zero-temperature
condensate with the same value of $N_0$. This is consistent with
earlier calculations~\cite{SK,PG} of the speed of sound in a homogeneous
Bose--condensed gas, which found that its temperature dependence
was effectively a condensate--density dependence.  We discuss
this result In a broader context in a separate paper~\cite{TWO_GAS},
in which we show that HFB-Popov results can be reproduced
quantititatively by a much simpler
``two--gas'' model: the condensate
gas, which is described by the
zero--temperature GP equation; and the
thermal cloud, which is
described as an ideal Bose gas in an effective potential
created by the condensate.  This effective
potential repels the thermal gas from the
condensate gas, which results in the essential
independence of temperature of all
condensate properties
except $N_0$.
Application of simple quantum
statistical mechanics to this model can generate the
full phase diagram of Fig.\ \ref{fig2} directly.

As Fig.~\ref{fig1} clearly shows, the HFB-Popov solutions reproduce
the experimental results quite well when $T \le 0.65~T_0$, but fail
at higher temperatures. The HFB-Popov formalism is biased
toward description of the condensate, as it represents
the condensate excitations
as taking place in a static thermal cloud.
This results in at least one minor
failure of the approach, which
is weakly visible in Fig. \ref{fig3} as a deviation of the $m=1$ mode
frequency from unity near $T_0$, in
violation of the generalized Kohn theorem for parabolic
confinement\cite{HUTCHINSON,GKT}.  This mode should
correspond to a rigid oscillation of the complete $N$-atom
system, and the deviation of its frequency
from unity results from the HFB-Popov
approximation holding the thermal component
fixed and allowing
only the condensate to oscillate.
In experiments of the type discussed here, however,
the thermal component and condensate must both
be driven by the modulation of their common
confining potential.
For other, non-rigid, oscillations we may thus also
expect thermal and
condensate modes to be coupled in general. Thus,
HFB-Popov frequencies will only
correspond to the experimental values if the condensate
response to mechanical disturbance
does not induce modulations of the thermal density,
so that there is no back-action of the thermal cloud on the
condensate motion. A more general theory that accounts for
such condensate-cloud interactions has recently been outlined
\cite{PROUKAKIS1}, but remains to be implemented.
The self-consistent inclusion of pair terms,
which are negelected in the Popov approximation,
may also
be important for capturing multiple-collision effects in the theory
of trapped BECs \cite{PROUKAKIS2}. The effect of such terms
can be very marked close to the transition temperature \cite{STOOF2}.

In conclusion, we have delineated the region of validity of the Popov
version of finite temperature HFB theory by comparing it directly
with the results of recent experiments.  Good agreement is
obtained for condensate fractions from unity down to about
0.5, so the HFB-Popov is apparently correctly describing
finite--temperature phenomena in a nontrivial regime.
This comparison confirms
the critical role of evaporatively cooled gases in establishing
proper finite-temperature field theories of Bose-Einstein condensation,
and shows that there is still work
needed to establish satisfactory
agreement between
theory and experiment
for cases of small condensate
fraction.

We thank the JILA group for providing us access to their experimental
data.  This work was supported in part by the U.\ S.\ National Science
Foundation under grants PHY-9601261 and PHY-9612728, the U.\ S.\ Office
of Naval Research, and the U.\ K.\ Engineering and Physical Sciences
Research Council.

\begin{references}

\bibitem[*]{NIST} Also at Physics Laboratory,
National Institute of Standards and Technology,
Technology Administration,
U.\ S.\ Department of Commerce,
Gaithersburg, MD 20899.

\bibitem{BEC!}  M.H.\ Anderson {\em et al.}, Science {\bf 269}, 198 (1995).

\bibitem{HULET} C.C.\ Bradley, C.A.\ Sackett, J.J.\ Tollett, and
R.G.\ Hulet, \prl {\bf 75}, 1687 (1995).

\bibitem{KETTERLE} K.B.\ Davis {\em et al.}, \prl {\bf 75}, 3969 (1995).

\bibitem{EXP_BEC_PROPERTIES} J.R.\ Ensher {\em et al.}, \prl {\bf 78},
764 (1997); D.S.\ Jin {\em et al.}, \prl {\bf 77}, 420 (1996);
M.-O.\ Mewes {\em et al.}, \prl {\bf 77}, 988 (1996).

\bibitem{POPOV} V.N.\ Popov, {\em Functional Integrals
and Collective Modes} (Cambridge University Press, New York, 1987),
Ch.\ 6.

\bibitem{T0_EXC_EXP} D.S.\ Jin {\em et al.}, \prl {\bf 77}, 420 (1996);
M.-O.\ Mewes {\em et al.}, \prl {\bf 77}, 992 (1996).

\bibitem{T0_EXC_THY} M.\ Edwards {\em et al.}, \prl {\bf 77}, 1671
(1996); S.\ Stringari, \prl {\bf 77}, 2360 (1996).

\bibitem{Popov_compare} A.\ Minguzzi, S.\ Conti, and M.\ P.\ Tosi,
J. Phys.: Condens. Matter, vol. 9, L33 (1997); S.\ Giorgini, L.\ P.\
Pitaevskii, and S.\ Stringari, \pra 54, 4633 (1996); S.\ Giorgini,
L.P. Pitaevskii, and S. Stringari, "Thermodynamics of a trapped,
Bose--condensed gas", unpublished (cond--mat/9704014).


\bibitem{GRIFFIN} A.\ Griffin, \prb {\bf 53}, 9341 (1996).

\bibitem{FETTER} A.L.\ Fetter, Ann.\ Phys.\ (NY) {\bf 70}, 1671 (1972).

\bibitem{Heinzen} D.\ Heinzen, private communication.






\bibitem{HUTCHINSON} D.A.W.\ Hutchinson, E.\ Zaremba, and A.\ Griffin,
\prl {\bf 78}, 1842 (1996).

\bibitem{T_DEPENDENT_EXC} D.S.\ Jin {\em et al.}, \prl {\bf 78}, 764
(1997).

\bibitem{EDCRB96} M.\ Edwards {\em et al.}, \pra {\bf 53}, R1950 (1996);
J.\ Res.\ Nat.\ Inst.\ Stand.\ Technol.\ {\bf 101}, 553 (1996).

\bibitem{SK} P.\ Sz\'epfalusy and I. Kondor, Ann. Phys. (N.Y.)
{\bf 82}, 1 (1974).

\bibitem{PG} S.\ H.\ Payne and A.\ Griffin, \prb {\bf 32}, 7199 (1985).

\bibitem{TWO_GAS} R.J.\ Dodd, K. Burnett, M. Edwards,
and C. W. Clark, \pra (submitted).

\bibitem{GKT} See J.F.\ Dobson, \prl {\bf 73},
2244 (1994); and references therein.

\bibitem{PROUKAKIS1} N.P.\ Proukakis and K.\ Burnett,
J.\ Res.\ Nat.\ Inst.\ Stand.\ Technol.\ {\bf 101}, 457 (1996).

\bibitem{PROUKAKIS2} N.P.\ Proukakis, K.\ Burnett, and
H.T.C.\ Stoof, preprint (1997).

\bibitem{STOOF2} M.\ Bijlsma and H.T.C.\ Stoof,
\pra {\bf 55}, 498 (1997).

\end{references}
\end{document}